\documentclass[pra,onecolumn, superscriptaddress, preprint]{revtex4}
\usepackage{amstext}
\usepackage{amsmath}
\usepackage{amssymb}

\newcommand{\ud}{\mathrm{d}}
\newcommand{\bra}[1]{\langle #1|}
\newcommand{\ket}[1]{| #1\rangle}
\newcommand{\vect}[1]{\boldsymbol{#1}}
\newcommand{\eq}[1]{Eq.~\eqref{#1}}

\newcommand{\stn}[1]{Sec.~\ref{#1}}
\newcommand{\be}{\begin{equation}}
\newcommand{\ee}{\end{equation}}
\newcommand{\ba}{\begin{align}}
\newcommand{\ea}{\end{align}}
\newcommand{\ti}[1]{\text{#1}}
\newcommand{\mc}[1]{\mathcal{#1}}
\newcommand{\w}{\omega}

\begin{document}

\title{Minimum requirements for laser-induced
symmetry breaking in quantum and classical  mechanics}
\author{Ignacio Franco and Paul Brumer\\
Chemical Physics Theory Group, Department of Chemistry, and \\
Center for Quantum Information and Quantum Control,\\ University
of Toronto, Toronto, Ontario  M5S 3H6, Canada.}

\date{\today}
\vspace{0.5in}
\begin{abstract}
Necessary conditions for generating  phase controllable asymmetry in spatially
symmetric systems using lasers  are identified and are shown to be identical in
quantum and classical mechanics.  First, by studying the exact dynamics of
harmonic systems in the presence of an arbitrary radiation field, it is demonstrated that
anharmonicities in the system's potential are a necessary requirement for phase controllability.
Then, by analyzing the space-time symmetries of the laser-driven
Liouville dynamics for classical and quantum systems, a common set of
temporal symmetries for the driving field that need to be violated to induce transport are identified. The conditions apply to continuous wave lasers and to symmetry breaking effects that do not rely on the control of the absolute phase of the field.
Known examples of laser fields that can induce transport in symmetric systems are
seen to be particular cases of these symmetry constraints.
\end{abstract}

\maketitle

\section{Introduction}

Recent years have witnessed the birth and rapid development of
the coherent control field ~\cite{paul,ricebook,bergmann,ricerev,dantusrev,nuernbergerrev}, in
which the coherence properties of applied laser fields are
employed to steer a given  quantum dynamical process in a desired
direction.  Of the different control schemes that have so far
been developed, there is a general class that has the ability to
induce phase controllable transport in spatially symmetric
systems without introducing a bias voltage in the potential, a
phenomenon that is referred to as laser-induced symmetry
breaking.

This symmetry breaking effect is typically achieved by driving
the system with  AC fields with frequency components $n\w$ and
$m\w$, where $n$ and $m$ are integers of different
parity~\cite{paul}. The nonlinear response of the system to such
fields results in  net dipoles or currents whose magnitude and
sign can be manipulated by varying the relative phase between the
two frequency components of the radiation~\cite{francoprl}. For
the popular case of $n=1$ and $m=2$ the rectification effect
first appears in the third order response of the system to the
incident radiation. At this order the system mixes the
frequencies and harmonics of the  field,  generating a
phase-controllable zero-harmonic (DC) component in the response.

Laser-induced symmetry breaking has been  demonstrated in a wide
variety of systems ranging from atoms to solid state samples.
Experimentally it has been implemented for generating anisotropy
in atomic photoionization~\cite{yin_1vs2}, symmetry breaking effects in
molecular photodissociation~\cite{sheehy_1vs2} (see also Ref.~\onlinecite{charron_1vs2}), photocurrents in
quantum wells~\cite{dupont_1vs2} and intrinsic
semiconductors~\cite{hache_1vs2}, as well as directed diffusion
in symmetric optical lattices~\cite{schiavoni}. Theoretically, it
has been studied for generating transport in
doped~\cite{paul_1vs2} and bulk semiconductors through
interband~\cite{atanasov_1vs2} and intraband~\cite{pronin_1vs2}
excitations, in graphene and carbon nanotubes~\cite{mele_1vs2},
and molecular wires~\cite{prlwire, lehmann_1vs2}, among others.
The scenario  is of interest since, with current laser
technology, it can be employed to generate transport on a
femtosecond timescale.

An  interesting feature of this laser control scenario is that it
has  both a quantum~\cite{paul_1vs2, hanggi_harmonic_mixing,
dupont_1vs2, hache_1vs2} and a classical~\cite{flach, denisov}
manifestation. Further, the two versions of the effect correspond
to the same physical phenomenon~\cite{francoprl},  arising from
the nonlinear response of material systems to symmetry breaking
radiation fields. In this contribution, we isolate minimum
conditions on the driving field and on the system that is being
driven that are necessary for the symmetry breaking effect
to occur in  quantum and classical mechanics. As shown, the
minimum requirements in both cases are identical and, further,
the effect can be accounted for in both mechanics through a
single symmetry analysis of the equations of motion.

Specifically, in \stn{sec:harmonic} we demonstrate   that laser
rectification can only occur in systems with anharmonic
potentials.   Subsequently, in \stn{sec:symmetry}  temporal
symmetries of the field  that need to be violated to induce
transport  are isolated. This is done  by studying the space-time
symmetries  of the  Liouville equations of motion for
laser-driven quantum and classical systems, and by isolating
symmetries of the field that rule out any nonzero average
currents or dipoles at asymptotic times. In doing so we
considerably extend a previous analysis~\cite{flach} that
identified conditions on the field  necessary for
laser-rectification in classical ergodic systems.  It applies to
both quantum and classical systems and makes no assumption of
ergodicity.

\section{Conditions on the system}
\label{sec:harmonic}

Consider first the exact solution for the dynamics of a harmonic
oscillator in the presence of an arbitrary space-homogeneous
radiation field $E(t)$. The Hamiltonian of the system reads
\begin{equation}
H(x,p) = \frac{p^2}{2m} + \frac{1}{2} m \omega_0^2 x^2 - q E(t) x,
\end{equation}
where $x$ and $p$ are the position and momentum of the particle
of mass $m$ and charge $q$, and $\w_0$ is the natural frequency
of the oscillator.  Symmetry breaking here would correspond to
the production of a net dipole moment.  In the quantum case,  the
dynamics of the position  $\hat{x}_\ti{H}(t)$ and momentum
operators  $\hat{p}_\ti{H}(t)$ in Heisenberg picture is dictated
by the Heisenberg equations of motion
\begin{subequations}
\label{eq:qset}
\begin{align}
\frac{\ud \hat{x}_\ti{H}(t)}{\ud t}  = \frac{1}{i \hbar}[\hat{x}_\ti{H}(t), \hat{H}_\ti{H}(t)] &  = \frac{1}{m} \hat{p}_\ti{H}(t), \\
 \frac{\ud \hat{p}_\ti{H}(t)}{\ud t}  = \frac{1}{i \hbar}[\hat{p}_\ti{H}(t), \hat{H}_\ti{H}(t)] &  =  -m \omega_0^2 \hat{x}_\ti{H}(t) + q E(t),
\end{align}
 \end{subequations}
where $\hat{H}_\ti{H}$ is  the Hamiltonian operator in Heisenberg
picture and $[\hat{f}, \hat{H}_\ti{H}]= \hat{f}\hat{H}_\ti{H} -
\hat{H}_\ti{H} \hat{f}$ for any operator $\hat{f}$. In the
classical case, the  position $x(t)$ and momentum $p(t)$
variables obey Hamilton's equations
\begin{subequations}
\label{eq:cset}
\begin{align}
\frac{\ud x(t)}{\ud t}  = \left\{x(t), H(t) \right\} &  = \frac{1}{m} p(t), \\
 \frac{\ud p(t)}{\ud t}  = \left\{p(t), H(t)\right\} &  =  -m \omega_0^2 x(t) + q E(t),
\end{align}
 \end{subequations}
where $\{f, H \}= \frac{\partial f}{\partial x}\frac{\partial
H}{\partial p} - \frac{\partial f}{\partial p}\frac{\partial
H}{\partial q}$ is the Poisson bracket. The difference between
Eqs.~\eqref{eq:qset} and~\eqref{eq:cset} is that the former is a
differential equation for operators, with operator initial
conditions  $\hat{x}_\ti{H}(0)=\hat{x}$ and
$\hat{p}_\ti{H}(0)=\hat{p}$, while the latter is an equation for
functions.

These two  sets of equations can be solved exactly using  Laplace
transforms. In fact, for a general external field of the form
\begin{equation}
\label{eq:field2}
E(t)= \int_{-\infty}^\infty  \, \ud \omega \,\epsilon({\omega}) e^{i\omega t},
\end{equation}
the usual procedure~\cite{arfken} yields:
\begin{multline}
\label{eq:quantum_pos}
\hat{x}_\ti{H}(t) =
\hat{x}_\ti{H}(0) \cos(\omega_0 t) +
\frac{\hat{p}_\ti{H}(0)}{m\omega_0} \sin(\omega_0 t) \\
 +
 \int_{-\infty}^\infty \ud\omega\,  \frac{q \epsilon(\omega)}{m\omega_0}
\frac{i\omega \sin({{\omega }_0}t) +
    {{\omega }_0}\cos({{\omega }_0}t) - {{\omega}_0}
    e^{i \omega t}}
    {{\omega }^2 - {{{\omega }_0}}^2};
\end{multline}
\begin{multline}
\label{eq:classical_pos}
x(t) = x(0) \cos(\omega_0 t) + \frac{p(0)}{m\omega_0} \sin(\omega_0 t) \\
+  \int_{-\infty}^\infty \ud\omega\, \frac{q \epsilon(\omega)}{m\omega_0}
\frac{i\omega \sin({{\omega }_0}t) +
    {{\omega }_0}\cos({{\omega }_0}t) - {{\omega}_0}
    e^{i \omega t}}
    {{\omega }^2 - {{{\omega }_0}}^2}.
\end{multline}
The first  two terms in Eqs.~\eqref{eq:quantum_pos}
and~\eqref{eq:classical_pos} describe the field-free evolution of
the oscillator, while the third one characterizes  the influence
of $E(t)$ on the dynamics.

Note that a  driven harmonic system can only oscillate at its
natural frequency $\omega_0$ and at the frequency of the field
$\omega$.  That is, there are no frequency components of the
dipole that oscillate at  multiples or combinations of the
frequencies of the field. Hence, if $E(t)$  is unbiased
($\epsilon(0) =0$)    no net dipole can be  induced. Thus, we
conclude that a necessary requirement for symmetry breaking in
quantum and classical mechanics is that the potential of the
system is anharmonic.  As seen below, the anharmonicities permit
the nonlinear response of the system to the incident radiation
that mixes the frequencies and harmonics of the field and, for a
certain class of radiation sources isolated below,  can lead to
the generation of a phase-controllable zero-harmonic (dc)
component in the response.

\section{Conditions on the field}
\label{sec:symmetry}

We now isolate those temporal symmetries of the field that need
to be violated to induce transport in both quantum and classical
mechanics.  To do so we consider a symmetric one-dimensional
system composed of  $N$ charged particles coupled to an
external   field $E(t)$  in  the dipole approximation. This is
done without loss of  generality  since the polarization of the
field effectively defines an axis  along which symmetry breaking
can arise. The system's Hamiltonian is then:
\begin{equation}
\label{eq:hamgen}
H = \sum_{j=1}^N \frac{p_j^2}{2m_j} + V(\vect{x}) -\sum_{j=1}^N q_j x_j  E(t+ \alpha\tfrac{T}{2\pi}),
\end{equation}
where $x_j$, $p_j$, $m_j$ and $q_j$ denote the coordinate,
momenta, mass and charge of the $j$-th particle and $\vect{x}
\equiv (x_1, x_2, \cdots, x_N)$. The systems of interest have a
potential  $V(\vect{x})$ that is invariant under  coordinate
inversion  $V(-\vect{x}) = V(\vect{x})$, and the driving field
$E(t+ \alpha\frac{T}{2\pi})$  is an arbitrary time-periodic
zero-mean function, with period $T$ and global phase $\alpha$.

In order to keep a close analogy  between the quantum and
classical case we frame this analysis  in phase space and adopt
the Wigner representation of quantum mechanics~\cite{ tatarskii,
wigner}. In it, the state of the quantum system is described by
the Wigner distribution function
$\rho_{\ti{W}}(\vect{x},\vect{p},t)$,  which constitutes a map of
the system density matrix $\hat{\rho}$ in the phase space of
 position $\vect{x}$ and momentum $\vect{p}$ variables.
For $N$-particle one-dimensional systems it is defined
by~\cite{wigner}
\be
\label{eq:wigner}
\rho_{\ti{W}}(\vect{x},\vect{p},t) =
           \left(\frac{1}{2\pi\hbar}\right)^N
\int_{-\infty}^{\infty}\cdots\int_{-\infty}^{\infty}
\ud u_1 \cdots \ud u_N  \,
 e^{ \frac{i}{\hbar} \vect{p}\cdot \vect{u} }
\bra{\vect{x} -\vect{u}/2} \hat{\rho}(t)\ket{\vect{x} + \vect{u}/2},
\ee
where  $\ket{\vect{x}}\equiv \ket{x_1}\ket{x_2}\cdots \ket{x_N}$ and $\vect{p} \cdot \vect{u} = \sum_{i=1}^N p_i u_i$.    In this way  the quantum or classical Liouville evolution can be expressed as
\begin{subequations}
\label{eq:dynamics}
\be
\label{eq:eqdynamics}
\mathcal{D}_{\beta} \rho_\beta(\vect{x}, \vect{p}, t)=0, \ee
where the label $\beta$ indicates  either classical
($\beta=\ti{c}$) or quantum ($\beta=\ti{W}$), with
$\rho_{\ti{c}}(\vect{x}, \vect{p},t)$  denoting the classical
phase space density. For the Hamiltonian in \eq{eq:hamgen}, the
operator $\mathcal{D}_\beta$  determining the  dynamics  is given
by~\cite{wigner, serimaa}
\begin{eqnarray}
\label{eq:classicaldynamics}
\mathcal{D}_{\text{c}} & = &
\frac{\partial}{\partial t} -
  \sum_{j=1}^{N}  \left[-\frac{p_j}{m_j} \frac{\partial}{\partial x_j}
 +  \left( \frac{\partial V(\vect{x})}{\partial x_j} -q_j E(t + \alpha\tfrac{T}{2\pi})\right)
         \frac{\partial}{\partial p_j}\right], \\
\label{eq:wignerdynamics}
\mathcal{D}_{\ti{W}} & = &
\mathcal{D}_{\ti{c}} - \mspace{-25.0mu}
\sum_{\substack{\lambda_1,\ldots,\lambda_N \\
       \lambda_1+\cdots+\lambda_N=3,5,\ldots}} \mspace{-25.0mu}
   \frac{\left(i \hbar/2\right)^{\lambda_1+\cdots+\lambda_N-1}}
              {\lambda_1! \cdots \lambda_N!}
       \frac{\partial^{\lambda_1+\cdots+\lambda_N }V(\vect{x})}
            {\partial x_1^{\lambda_1}\cdots{\partial x_N^{\lambda_N}}}
       \frac{\partial^{\lambda_1+\cdots+\lambda_N}}
            {\partial p_1^{\lambda_1}\cdots{\partial p_N^{\lambda_N}}},
\end{eqnarray}
\end{subequations}
where the last summation in $\mc{D}_\ti{W}$ runs over all
positive integer  values of $\lambda_1,\ldots,\lambda_N$ for
which the sum $\lambda_1+\lambda_2+\cdots+\lambda_N$ is odd and
greater than one.  In phase space the formal structure of the
quantum and classical evolution is remarkably
similar~\cite{wilkie1, wilkie2}. In the limit $\hbar\to 0$ the
second term in \eq{eq:wignerdynamics} vanishes and the quantum
equation of motion reduces to the classical evolution
($\mathcal{D}_\text{w}\to \mathcal{D}_\text{c}$). Note that
Eqs.~\eqref{eq:wigner} and~\eqref{eq:dynamics} are fully
consistent with the Hamiltonian in \eq{eq:hamgen}.  However, when
the radiation-matter interaction term in the Hamiltonian goes
beyond the dipole approximation both of them need to be modified
in order to ensure gauge invariance~\cite{serimaa}.

In the absence of an external field  the equations of motion
[\eq{eq:dynamics}]  are invariant under reflection ($\vect{x}\to
-\vect{x}$, $\vect{p}\to -\vect{p}$). Hence, if the system is
initially prepared with a given phase-space symmetry this initial
symmetry is preserved at all times during the subsequent
dynamics.   Symmetry breaking is achieved  by coupling  the
system to $E(t)$. However, if $E(t)$ has a zero temporal mean (AC
field) then not every $E(t)$ will generate transport.  As shown
below, by properly lowering the temporal symmetry of $E(t)$ it is
possible to induce rectification in the response.  Furthermore,
the resulting symmetry constraints on $E(t)$  are identical for
the classical and quantum case. As will become evident, this is
a  consequence of the important fact that the quantum corrections
in $\mc{D}_\ti{W}$   have the same symmetry properties as
$\frac{\partial V(\vect{x})}{\partial x_j}
\frac{\partial}{\partial p_j}$ under inversion of position and
momentum coordinates.

We focus on rectification effects that survive time averaging
and that are independent of the global phase $\alpha$ of the
laser beam. Typically, symmetry breaking effects that depend on
$\alpha$ are difficult to control (although not
impossible~\cite{kling})  since this requires an experimental
setup that both locks the absolute phase of the laser and has
control over the center of mass  motion with respect to the
laboratory frame.  Hence, the quantities of interest  are the
mean position and momentum averaged over time and over $\alpha$:
\begin{subequations}
\label{eq:mean}
\begin{align}
\langle \overline{\overline{\vect{x}}} \rangle_{\beta} &
= \lim_{\tau\to\infty}
\int_{-\tau/2}^{\tau/2} \frac{\ud t}{\tau}
 \int_{0}^{2\pi}
\frac{\ud\alpha}{2\pi} \text{Tr}(\vect{x}   \rho_{\beta}(\vect{x},\vect{p},t)); \\
\langle \overline{\overline{\vect{p}}} \rangle_\beta &
= \lim_{\tau\to\infty}
\int_{-\tau/2}^{\tau/2} \frac{\ud t}{\tau}
 \int_{0}^{2\pi}
\frac{\ud\alpha}{2\pi}   \text{Tr}(\vect{p}\rho_{\beta}(\vect{x},\vect{p},t));
\end{align}
\end{subequations}
where the double overbar indicates this kind of averaging.  Here
the notation $\langle \cdots \rangle_{\beta}$ denotes the
classical ensemble average ($\beta=c$) or quantum expectation
value ($\beta=\ti{W}$), and  the trace is an integration over the
$2N$-dimensional phase-space $(\vect{x}, \vect{p})$.    When the
symmetry of the system is not broken,  both $\langle
\overline{\overline{\vect{x}}} \rangle_{\beta}$ and $\langle
\overline{\overline{\vect{p}}} \rangle_{\beta}$ are zero. Below
we determine symmetries of the field and of the initial condition
that guarantee that this is indeed the case. When these
symmetries are violated a net dipole or current is expected to
appear.

The fact that we are only interested in $\alpha$-independent
properties eliminates the necessity to  invoke ergodicity in the
analysis.  The average over $\alpha$ is sufficient to obviate any
initial-time preparation effects, which is the main role of the
ergodicity assumption in the purely classical analysis of
Ref.~\cite{flach}.

We now tabulate the symmetries of the field   that will be
relevant for our purposes.  The field may change sign every half
a period $T$,
\begin{subequations}
\label{eq:sym}
\begin{align}
E(t + T/2) = &-  E(t);\label{eq:sym1}\\
\intertext{or be symmetric or antisymmetric with respect to some time $t'$}
E(t - t') =  &+ E(-(t-t')); \label{eq:sym2}\\
E(t - t') = &-  E(-(t-t')). \label{eq:sym3}
\end{align}
\end{subequations}
Each of the  symmetries  in \eq{eq:sym}   leads to a
transformation that leaves the equations of motion invariant
while  changing the sign of the position and/or momentum
variables. They are identical  for the quantum and classical
case. For example, if $E(t)$ satisfies \eq{eq:sym1}, then
$\mathcal{D}_{\beta}$  is invariant under $\mathcal{T}_1$ defined
as:
\begin{subequations}
\label{eq:trans}
\begin{align}
\label{eq:trans1}
 \mathcal{T}_1&: \quad    t  \to  t+T/2;\quad  \vect{x}  \to -\vect{x}; \quad  \vect{p}  \to -\vect{p}; \\
\intertext{where we have taken into account that under inversion of position and momenta,
$\frac{\partial^{\lambda_1+\cdots+\lambda_N }}
      {\partial x_1^{\lambda_1}\cdots{\partial x_N^{\lambda_N}}}
\to -
 \frac{\partial^{\lambda_1+\cdots+\lambda_N }}
      {\partial x_1^{\lambda_1}\cdots{\partial x_N^{\lambda_N}}}$ and
$\frac{\partial^{\lambda_1+\cdots+\lambda_N }}
      {\partial p_1^{\lambda_1}\cdots{\partial p_N^{\lambda_N}}}
\to -
 \frac{\partial^{\lambda_1+\cdots+\lambda_N }}
      {\partial p_1^{\lambda_1}\cdots{\partial p_N^{\lambda_N}}}$ since  $\lambda_1 + \lambda_2 + \cdots +\lambda_N$ in \eq{eq:wignerdynamics} is odd.
Similarly, if $E(t)$ satisfies \eq{eq:sym2} [or \eq{eq:sym3}] then  $\mathcal{D}_{\beta}$ is invariant under $\mathcal{T}_2$ [or $\mathcal{T}_3$], where}
\label{eq:trans2}
  \mathcal{T}_2&:\, \quad  t-t'  \to  -(t-t'); \quad \vect{x}  \to \vect{x}; \quad \vect{p}  \to -\vect{p}; \\
\label{eq:trans3}
  \mathcal{T}_3&:\,\quad   t-t'  \to -(t-t'); \quad \vect{x}  \to -\vect{x};  \quad \vect{p}  \to \vect{p}.
\end{align}
\end{subequations}
Other temporal symmetries of the field  exist but play no role in this analysis since they do not lead to invariance transformations that change the sign of the position and/or momentum variables.

Now, given a solution  to \eq{eq:dynamics}, $\rho_\beta(\vect{x}, \vect{p}, t)$, if $\mc{D}_\beta$ is invariant under  $\mc{T}_\alpha$ one can generate another solution to the same equation  by applying $\mc{T}_\alpha$ to  $\rho_\beta(\vect{x}, \vect{p}, t)$. The new solutions  $\rho_{\beta}^{(\alpha)}(\vect{x}, \vect{p}, t) = \mathcal{T}_\alpha \rho_{\beta}(\vect{x}, \vect{p}, t)$ generated by the   invariance transformations  in \eq{eq:trans} are:
\begin{subequations}
\label{eq:transfrho}
\begin{eqnarray}
\label{eq:transfrho1}
\rho_{\beta}^{(1)}(\vect{x}, \vect{p}, t)   = &
\mathcal{T}_1 \rho_{\beta}(\vect{x}, \vect{p},t)
   = &    \rho_{\beta}(-\vect{x}, -\vect{p},t+T/2); \\
\rho_{\beta}^{(2)}(\vect{x}, \vect{p}, t)   = &  \mathcal{T}_2 \rho_{\beta}(\vect{x}, \vect{p}, t-t')  = & \rho_{\beta}(\vect{x}, -\vect{p},-(t-t')); \\
\rho_{\beta}^{(3)}(\vect{x}, \vect{p}, t)  = & \mathcal{T}_3 \rho_{\beta}(\vect{x}, \vect{p},t-t')
  = & \rho_{\beta}(-\vect{x}, \vect{p},-(t-t')).
\end{eqnarray}
\end{subequations}
Further, if the original solution $\rho_{\beta}(\vect{x}, \vect{p}, t)$ predicts an average position $\langle \overline{\overline{\vect{x}}}\rangle_\beta$ and momenta    $\langle \overline{\overline{\vect{p}}}\rangle_\beta$, the transformed solutions $\rho_{\beta}^{(\alpha)}(\vect{x}, \vect{p}, t)$  will predict a mean position $\langle \overline{\overline{\vect{x}}}\rangle_\beta^{(\alpha)}$ and/or momenta $\langle \overline{\overline{\vect{p}}}\rangle_\beta^{(\alpha)}$ that has  the same magnitude but is opposite in sign:
\begin{subequations}
\label{eq:expectt}
\begin{eqnarray}
\label{eq:expect1}
 \langle \overline{\overline{\vect{x}}}\rangle_\beta^{(1)} = -
 \langle \overline{\overline{\vect{x}}}\rangle_\beta;
&\langle \overline{\overline{\vect{p}}}\rangle_\beta^{(1)} = -
 \langle \overline{\overline{\vect{p}}}\rangle_\beta; \\
\label{eq:expect2}
\langle \overline{\overline{\vect{x}}}\rangle_\beta^{(2)} = +
\langle \overline{\overline{\vect{x}}}\rangle_\beta;
&
\langle \overline{\overline{\vect{p}}}\rangle_\beta^{(2)} = -
\langle \overline{\overline{\vect{p}}}\rangle_\beta; \\
\label{eq:expect3}
\langle  \overline{\overline{\vect{x}}}\rangle_\beta^{(3)} = -
\langle  \overline{\overline{\vect{x}}}\rangle_\beta;
& \langle \overline{\overline{\vect{p}}}\rangle_\beta^{(3)} = +
\langle \overline{\overline{\vect{p}}}\rangle_\beta.
\end{eqnarray}
\end{subequations}

The argument is completed by showing that the average position
and momenta  predicted by  $\rho_{\beta}(\vect{x}, \vect{p}, t)$
and $\rho_{\beta}^{(\alpha)}(\vect{x}, \vect{p}, t)$  coincide.
If this is the case, it follows  from \eq{eq:expectt} that
symmetry breaking cannot occur.   For this we exploit the
possible symmetries of the initial state:
\begin{subequations}
\label{eq:initialsym}
\begin{align}
\label{eq:initialsym1}
\rho_{\beta}(\vect{x}, \vect{p}, t_0)  & =  \rho_{\beta}(-\vect{x}, -\vect{p}, t_0);\\
\label{eq:initialsym2}
\rho_{\beta}(\vect{x}, \vect{p}, t_0)  & =  \rho_{\beta}(\vect{x}, -\vect{p}, t_0); \\
\label{eq:initialsym3}
\rho_{\beta}(\vect{x}, \vect{p}, t_0)  & =  \rho_{\beta}(-\vect{x}, \vect{p}, t_0).
\end{align}
\end{subequations}
The first one corresponds to a state  with zero mean position and
momenta, while the second and third describe an initial state
with either zero mean momenta or zero mean position,
respectively.

 Consider the case in which the equations  of motion are $\mathcal{T}_1$ invariant.
 The distributions $\rho_{\beta}(\vect{x}, \vect{p}, t)$ and $\rho_{\beta}^{(1)}(\vect{x}, \vect{p}, t)$
  satisfy the same equation of motion but do not, in general, coincide.  However,
    if  the initial condition for the original solution  $\rho_{\beta}(\vect{x}, \vect{p}, t_0)$
     is invariant under reflection [\eq{eq:initialsym1}]  then
\be
\label{eq:inter1}
\rho_{\beta}^{(1)}(\vect{x}, \vect{p}, t_0 - T/2)=
\rho_{\beta}(-\vect{x}, -\vect{p}, t_0) =  \rho_{\beta}(\vect{x},
\vect{p}, t_0) = \rho_{\beta}^{(1)}(-\vect{x}, -\vect{p}, t_0 -
T/2). \ee That is, the original and transformed solutions  start
from the same initial distribution but at initial time they
experience a different value for the global phase of the field,
$E(t_0 +\alpha\frac{T}{2\pi})$ and
$E(t_0+(\alpha-\pi)\frac{T}{2\pi})=-E(t_0+ \alpha\frac{T}{2\pi})$
respectively. Since the averages in \eq{eq:mean} are independent
of $\alpha$,  they coincide  for the two solutions.  Hence,  no
rectification can be induced when the field satisfies
\eq{eq:sym1} and the initial condition satisfies
\eq{eq:initialsym1}.

The argument for the  two other cases is very similar.  If the
field satisfies \eq{eq:sym2} [or \eq{eq:sym3}], the equation of
motion is $\mc{T}_2$ (or $\mc{T}_3$) invariant. Even when the
original $\rho_{\beta}(\vect{x}, \vect{p}, t)$ and transformed
$\rho_{\beta}^{(2)}(\vect{x}, \vect{p}, t)$ [or
$\rho_{\beta}^{(3)}(\vect{x}, \vect{p}, t)$] solutions obey the
same equation of motion, they do not need to coincide. However,
if  the initial condition of the original solution satisfies the
symmetry in \eq{eq:initialsym2} [or \eq{eq:initialsym3}], then
\begin{align}
\rho_{\beta}^{(2)}(\vect{x}, \vect{p}, -t_0+t')&= \rho_{\beta}(\vect{x}, -\vect{p}, t_0) =  \rho_{\beta}(\vect{x}, \vect{p}, t_0) = \rho_{\beta}^{(2)}(\vect{x}, -\vect{p}, -t_0+t')\\
\rho_{\beta}^{(3)}(\vect{x}, \vect{p}, -t_0+t')&= \rho_{\beta}(-\vect{x}, \vect{p}, t_0) =  \rho_{\beta}(\vect{x}, \vect{p}, t_0) = \rho_{\beta}^{(3)}(-\vect{x}, \vect{p}, -t_0+t')
\end{align}
 The transformed solution has the same initial condition as
  the original one but as we had prepared the system a time $2t_0 - t'$ before.
  The difference between the two solutions is that they experience a different
  global laser phase at preparation time. Since we are not interested in effects
   that depend on the global laser phase, the mean position and momenta
   in \eq{eq:mean} for the original and transformed solution need to coincide.
    However, from \eq{eq:expect2} [or \eq{eq:expect3}] we conclude that this
    can only happen if $\langle \overline{\overline{\vect{p}}} \rangle_\beta = 0$
     [or $\langle \overline{\overline{\vect{x}}} \rangle_\beta = 0$].

In summary, for spatially symmetric \emph{classical or quantum}
systems  initially prepared in a symmetric state that satisfies
\eq{eq:initialsym},  net  transport using time-periodic external
fields with zero temporal mean can only be generated if the field
violates the temporal symmetries in \eq{eq:sym}.  Further,  any
symmetry breaking effect that may be achieved with a field that
satisfies \eq{eq:sym} is necessarily due to an effect that
depends on the global phase of the laser (cf. Ref.~\cite{kling}).

It is  natural to ask what kind of fields have sufficiently low
temporal  symmetry to induce net transport.   Monochromatic
sources satisfy all the symmetries in \eq{eq:sym} and, as a
consequence, cannot be used to induce symmetry breaking.
However, by adding a second frequency  component to a
monochromatic source it is possible to  lower the symmetry of the
field and induce symmetry breaking. For instance, a field like
\begin{equation}
\label{eq:field}
E(t) = \epsilon_{n\w} \cos(n \omega t + \phi_{n\omega}) +
       \epsilon_{m\w} \cos(m \omega t + \phi_{m\omega}),
\end{equation}
where $n$ and $m$ are coprime  integers so that $E(t)$ has a period $T= 2\pi/\omega$, satisfies \eq{eq:sym} only under special conditions. It satisfies \eqref{eq:sym1} only if $n$ and $m$ are odd. Thus, a field with $n=3$ and $m=1$, like the one used in the  1 vs. 3 photon control scenario~\cite{paul}, will not be symmetry breaking.
However,  a field with  $n=2$ and $m=1$,  like the one employed in the 1 vs. 2 scenario, does not  satisfy \eq{eq:sym} and is expected to induce  net dipoles and currents.  These dipoles and currents are phase-controllable since, by varying the relative phase between the two components of the beam, the $\w+2\w$ field may satisfy \eq{eq:sym2} or~\eqref{eq:sym3} and thus rule out the possibility of inducing currents or dipoles, respectively.  Explicitly, when  $\phi_{2\omega} - 2\phi_{\omega}= 0, \pm\pi, \pm 2\pi, \cdots$  an $\w+2\w$ field satisfies \eq{eq:sym2} and zero currents are expected.  Similarly, when  $\phi_{2\omega} - 2\phi_{\omega}= \pm\frac{\pi}{2}, \pm\frac{3\pi}{2}, \cdots$ it fulfills symmetry \eqref{eq:sym3}  and no dipoles can be induced. For all other choices of the phases an $\w+2\w$ field is expected to induce symmetry breaking.

\section{Conclusions}
\label{sec:conclusions}

In conclusion, we have shown that the minimum conditions for the
generation of phase controllable asymmetry in spatially symmetric
quantum and classical systems using time-periodic external fields
with zero temporal mean are identical: anharmonicities in the
system's potential are required as is a driving field that violates
the temporal symmetries in \eq{eq:sym}. These conditions refer to
symmetry breaking effects that do not rely on the control of the absolute 
phase of the field. The derived results provide necessary conditions for
the generation of asymmetry, applicable to all systems.
Additional sufficient conditions may be required, but these depend upon 
the specific system under consideration.  

Further, we have shown that both quantum and
classical versions of the symmetry breaking effect can be accounted for 
through a single space-time symmetry analysis of the equations of motion.
The anharmonicities in the  potential permit the nonlinear
response of the system to the incident radiation that, through
harmonic mixing, and for fields that violate \eq{eq:sym}, can lead
to the generation of a phase-controllable DC component in the
photoinduced dipoles or currents.

\section{Acknowledgments} This work was supported by NSERC Canada.

\bibliography{symmetry_analysis}

\begin{thebibliography}{29}
\expandafter\ifx\csname natexlab\endcsname\relax\def\natexlab#1{#1}\fi
\expandafter\ifx\csname bibnamefont\endcsname\relax
  \def\bibnamefont#1{#1}\fi
\expandafter\ifx\csname bibfnamefont\endcsname\relax
  \def\bibfnamefont#1{#1}\fi
\expandafter\ifx\csname citenamefont\endcsname\relax
  \def\citenamefont#1{#1}\fi
\expandafter\ifx\csname url\endcsname\relax
  \def\url#1{\texttt{#1}}\fi
\expandafter\ifx\csname urlprefix\endcsname\relax\def\urlprefix{URL }\fi
\providecommand{\bibinfo}[2]{#2}
\providecommand{\eprint}[2][]{\url{#2}}

\bibitem[{\citenamefont{Shapiro and Brumer}(2003)}]{paul}
\bibinfo{author}{\bibfnamefont{M.}~\bibnamefont{Shapiro}} \bibnamefont{and}
  \bibinfo{author}{\bibfnamefont{P.}~\bibnamefont{Brumer}},
  \emph{\bibinfo{title}{Principles of the Quantum Control of Molecular
  Processes}} (\bibinfo{publisher}{John Wiley \& Sons}, \bibinfo{address}{New
  York}, \bibinfo{year}{2003}).

\bibitem[{\citenamefont{Rice and Zhao}(2000)}]{ricebook}
\bibinfo{author}{\bibfnamefont{S.~A.} \bibnamefont{Rice}} \bibnamefont{and}
  \bibinfo{author}{\bibfnamefont{M.}~\bibnamefont{Zhao}},
  \emph{\bibinfo{title}{Optical Control of Molecular Dynamics}}
  (\bibinfo{publisher}{John Wiley \& Sons}, \bibinfo{address}{New York},
  \bibinfo{year}{2000}).

\bibitem[{\citenamefont{Bergmann et~al.}(1998)\citenamefont{Bergmann, Theuer,
  and Shore}}]{bergmann}
\bibinfo{author}{\bibfnamefont{K.}~\bibnamefont{Bergmann}},
  \bibinfo{author}{\bibfnamefont{H.}~\bibnamefont{Theuer}}, \bibnamefont{and}
  \bibinfo{author}{\bibfnamefont{B.~W.} \bibnamefont{Shore}},
  \bibinfo{journal}{Rev. Mod. Phys.} \textbf{\bibinfo{volume}{70}},
  \bibinfo{pages}{1003} (\bibinfo{year}{1998}).

\bibitem[{\citenamefont{Rice}(2001)}]{ricerev}
\bibinfo{author}{\bibfnamefont{S.~A.} \bibnamefont{Rice}},
  \bibinfo{journal}{Nature} \textbf{\bibinfo{volume}{409}},
  \bibinfo{pages}{422} (\bibinfo{year}{2001}).

\bibitem[{\citenamefont{Dantus and Lozovoy}(2004)}]{dantusrev}
\bibinfo{author}{\bibfnamefont{M.}~\bibnamefont{Dantus}} \bibnamefont{and}
  \bibinfo{author}{\bibfnamefont{V.~V.} \bibnamefont{Lozovoy}},
  \bibinfo{journal}{Chem. Rev.} \textbf{\bibinfo{volume}{104}},
  \bibinfo{pages}{1813} (\bibinfo{year}{2004}).

\bibitem[{\citenamefont{Nuernberger et~al.}(2007)\citenamefont{Nuernberger,
  Vogt, Brixner, and Gerber}}]{nuernbergerrev}
\bibinfo{author}{\bibfnamefont{P.}~\bibnamefont{Nuernberger}},
  \bibinfo{author}{\bibfnamefont{G.}~\bibnamefont{Vogt}},
  \bibinfo{author}{\bibfnamefont{T.}~\bibnamefont{Brixner}}, \bibnamefont{and}
  \bibinfo{author}{\bibfnamefont{G.}~\bibnamefont{Gerber}},
  \bibinfo{journal}{Phys. Chem. Chem. Phys.} \textbf{\bibinfo{volume}{9}},
  \bibinfo{pages}{2470} (\bibinfo{year}{2007}).

\bibitem[{\citenamefont{Franco and Brumer}(2006)}]{francoprl}
\bibinfo{author}{\bibfnamefont{I.}~\bibnamefont{Franco}} \bibnamefont{and}
  \bibinfo{author}{\bibfnamefont{P.}~\bibnamefont{Brumer}},
  \bibinfo{journal}{Phys. Rev. Lett.} \textbf{\bibinfo{volume}{97}},
  \bibinfo{pages}{040402} (\bibinfo{year}{2006}).

\bibitem[{\citenamefont{Yin et~al.}(1992)\citenamefont{Yin, Chen, Elliott, and
  Smith}}]{yin_1vs2}
\bibinfo{author}{\bibfnamefont{Y.~Y.} \bibnamefont{Yin}},
  \bibinfo{author}{\bibfnamefont{C.}~\bibnamefont{Chen}},
  \bibinfo{author}{\bibfnamefont{D.~S.} \bibnamefont{Elliott}},
  \bibnamefont{and} \bibinfo{author}{\bibfnamefont{A.~V.} \bibnamefont{Smith}},
  \bibinfo{journal}{Phys. Rev. Lett.} \textbf{\bibinfo{volume}{69}},
  \bibinfo{pages}{2353} (\bibinfo{year}{1992}).

\bibitem[{\citenamefont{Sheehy et~al.}(1995)\citenamefont{Sheehy, Walker, and
  DiMauro}}]{sheehy_1vs2}
\bibinfo{author}{\bibfnamefont{B.}~\bibnamefont{Sheehy}},
  \bibinfo{author}{\bibfnamefont{B.}~\bibnamefont{Walker}}, \bibnamefont{and}
  \bibinfo{author}{\bibfnamefont{L.~F.} \bibnamefont{DiMauro}},
  \bibinfo{journal}{Phys. Rev. Lett.} \textbf{\bibinfo{volume}{74}},
  \bibinfo{pages}{4799} (\bibinfo{year}{1995}).

\bibitem[{\citenamefont{Charron et~al.}(1995)\citenamefont{Charron,
  Giusti-Suzor, and Mies}}]{charron_1vs2}
\bibinfo{author}{\bibfnamefont{E.}~\bibnamefont{Charron}},
  \bibinfo{author}{\bibfnamefont{A.}~\bibnamefont{Giusti-Suzor}},
  \bibnamefont{and} \bibinfo{author}{\bibfnamefont{F.~H.} \bibnamefont{Mies}},
  \bibinfo{journal}{Phys. Rev. Lett.} \textbf{\bibinfo{volume}{75}},
  \bibinfo{pages}{2815} (\bibinfo{year}{1995}).

\bibitem[{\citenamefont{Dupont et~al.}(1995)\citenamefont{Dupont, Corkum, Liu,
  Buchanan, and Wasilewski}}]{dupont_1vs2}
\bibinfo{author}{\bibfnamefont{E.}~\bibnamefont{Dupont}},
  \bibinfo{author}{\bibfnamefont{P.~B.} \bibnamefont{Corkum}},
  \bibinfo{author}{\bibfnamefont{H.~C.} \bibnamefont{Liu}},
  \bibinfo{author}{\bibfnamefont{M.}~\bibnamefont{Buchanan}}, \bibnamefont{and}
  \bibinfo{author}{\bibfnamefont{Z.~R.} \bibnamefont{Wasilewski}},
  \bibinfo{journal}{Phys. Rev. Lett.} \textbf{\bibinfo{volume}{74}},
  \bibinfo{pages}{3596} (\bibinfo{year}{1995}).

\bibitem[{\citenamefont{Hach\'e et~al.}(1997)\citenamefont{Hach\'e, Kostoulas,
  Atanasov, Hughes, Sipe, and van Driel}}]{hache_1vs2}
\bibinfo{author}{\bibfnamefont{A.}~\bibnamefont{Hach\'e}},
  \bibinfo{author}{\bibfnamefont{Y.}~\bibnamefont{Kostoulas}},
  \bibinfo{author}{\bibfnamefont{R.}~\bibnamefont{Atanasov}},
  \bibinfo{author}{\bibfnamefont{J.~L.~P.} \bibnamefont{Hughes}},
  \bibinfo{author}{\bibfnamefont{J.~E.} \bibnamefont{Sipe}}, \bibnamefont{and}
  \bibinfo{author}{\bibfnamefont{H.~M.} \bibnamefont{van Driel}},
  \bibinfo{journal}{Phys. Rev. Lett.} \textbf{\bibinfo{volume}{78}},
  \bibinfo{pages}{306} (\bibinfo{year}{1997}).

\bibitem[{\citenamefont{Schiavoni et~al.}(2003)\citenamefont{Schiavoni,
  Sanchez-Palencia, Renzoni, and Grynberg}}]{schiavoni}
\bibinfo{author}{\bibfnamefont{M.}~\bibnamefont{Schiavoni}},
  \bibinfo{author}{\bibfnamefont{L.}~\bibnamefont{Sanchez-Palencia}},
  \bibinfo{author}{\bibfnamefont{F.}~\bibnamefont{Renzoni}}, \bibnamefont{and}
  \bibinfo{author}{\bibfnamefont{G.}~\bibnamefont{Grynberg}},
  \bibinfo{journal}{Phys. Rev. Lett.} \textbf{\bibinfo{volume}{90}},
  \bibinfo{pages}{094101} (\bibinfo{year}{2003}).

\bibitem[{\citenamefont{Kurizki et~al.}(1989)\citenamefont{Kurizki, Shapiro,
  and Brumer}}]{paul_1vs2}
\bibinfo{author}{\bibfnamefont{G.}~\bibnamefont{Kurizki}},
  \bibinfo{author}{\bibfnamefont{M.}~\bibnamefont{Shapiro}}, \bibnamefont{and}
  \bibinfo{author}{\bibfnamefont{P.}~\bibnamefont{Brumer}},
  \bibinfo{journal}{Phys. Rev. B} \textbf{\bibinfo{volume}{39}},
  \bibinfo{pages}{3435} (\bibinfo{year}{1989}).

\bibitem[{\citenamefont{Atanasov et~al.}(1996)\citenamefont{Atanasov, Hach\'e,
  Hughes, van Driel, and Sipe}}]{atanasov_1vs2}
\bibinfo{author}{\bibfnamefont{R.}~\bibnamefont{Atanasov}},
  \bibinfo{author}{\bibfnamefont{A.}~\bibnamefont{Hach\'e}},
  \bibinfo{author}{\bibfnamefont{J.~L.~P.} \bibnamefont{Hughes}},
  \bibinfo{author}{\bibfnamefont{H.~M.} \bibnamefont{van Driel}},
  \bibnamefont{and} \bibinfo{author}{\bibfnamefont{J.~E.} \bibnamefont{Sipe}},
  \bibinfo{journal}{Phys. Rev. Lett.} \textbf{\bibinfo{volume}{76}},
  \bibinfo{pages}{1703} (\bibinfo{year}{1996}).

\bibitem[{\citenamefont{Pronin and Bandrauk}(2004)}]{pronin_1vs2}
\bibinfo{author}{\bibfnamefont{K.~A.} \bibnamefont{Pronin}} \bibnamefont{and}
  \bibinfo{author}{\bibfnamefont{A.~D.} \bibnamefont{Bandrauk}},
  \bibinfo{journal}{Phys. Rev. B} \textbf{\bibinfo{volume}{69}},
  \bibinfo{pages}{195308} (\bibinfo{year}{2004}).

\bibitem[{\citenamefont{Mele et~al.}(2000)\citenamefont{Mele, Kr\'al, and
  Tom\'anek}}]{mele_1vs2}
\bibinfo{author}{\bibfnamefont{E.~J.} \bibnamefont{Mele}},
  \bibinfo{author}{\bibfnamefont{P.}~\bibnamefont{Kr\'al}}, \bibnamefont{and}
  \bibinfo{author}{\bibfnamefont{D.}~\bibnamefont{Tom\'anek}},
  \bibinfo{journal}{Phys. Rev. B} \textbf{\bibinfo{volume}{61}},
  \bibinfo{pages}{7669} (\bibinfo{year}{2000}).

\bibitem[{\citenamefont{Franco et~al.}(2007)\citenamefont{Franco, Shapiro, and
  Brumer}}]{prlwire}
\bibinfo{author}{\bibfnamefont{I.}~\bibnamefont{Franco}},
  \bibinfo{author}{\bibfnamefont{M.}~\bibnamefont{Shapiro}}, \bibnamefont{and}
  \bibinfo{author}{\bibfnamefont{P.}~\bibnamefont{Brumer}},
  \bibinfo{journal}{Phys. Rev. Lett.} \textbf{\bibinfo{volume}{99}},
  \bibinfo{pages}{126802} (\bibinfo{year}{2007}).

\bibitem[{\citenamefont{Lehmann et~al.}(2004)\citenamefont{Lehmann, Kohler,
  May, and Hanggi}}]{lehmann_1vs2}
\bibinfo{author}{\bibfnamefont{J.}~\bibnamefont{Lehmann}},
  \bibinfo{author}{\bibfnamefont{S.}~\bibnamefont{Kohler}},
  \bibinfo{author}{\bibfnamefont{V.}~\bibnamefont{May}}, \bibnamefont{and}
  \bibinfo{author}{\bibfnamefont{P.}~\bibnamefont{Hanggi}},
  \bibinfo{journal}{J. Chem. Phys.} \textbf{\bibinfo{volume}{121}},
  \bibinfo{pages}{2278} (\bibinfo{year}{2004}).

\bibitem[{\citenamefont{Goychuk and H\"anggi}(1998)}]{hanggi_harmonic_mixing}
\bibinfo{author}{\bibfnamefont{I.}~\bibnamefont{Goychuk}} \bibnamefont{and}
  \bibinfo{author}{\bibfnamefont{P.}~\bibnamefont{H\"anggi}},
  \bibinfo{journal}{Europhys. Lett.} \textbf{\bibinfo{volume}{43}},
  \bibinfo{pages}{503} (\bibinfo{year}{1998}).

\bibitem[{\citenamefont{Flach et~al.}(2000)\citenamefont{Flach, Yevtushenko,
  and Zolotaryuk}}]{flach}
\bibinfo{author}{\bibfnamefont{S.}~\bibnamefont{Flach}},
  \bibinfo{author}{\bibfnamefont{O.}~\bibnamefont{Yevtushenko}},
  \bibnamefont{and}
  \bibinfo{author}{\bibfnamefont{Y.}~\bibnamefont{Zolotaryuk}},
  \bibinfo{journal}{Phys. Rev. Lett.} \textbf{\bibinfo{volume}{84}},
  \bibinfo{pages}{2358} (\bibinfo{year}{2000}).

\bibitem[{\citenamefont{Denisov et~al.}(2002)\citenamefont{Denisov, Flach,
  Ovchinnikov, Yevtushenko, and Zolotaryuk}}]{denisov}
\bibinfo{author}{\bibfnamefont{S.}~\bibnamefont{Denisov}},
  \bibinfo{author}{\bibfnamefont{S.}~\bibnamefont{Flach}},
  \bibinfo{author}{\bibfnamefont{A.~A.} \bibnamefont{Ovchinnikov}},
  \bibinfo{author}{\bibfnamefont{O.}~\bibnamefont{Yevtushenko}},
  \bibnamefont{and}
  \bibinfo{author}{\bibfnamefont{Y.}~\bibnamefont{Zolotaryuk}},
  \bibinfo{journal}{Phys. Rev. E} \textbf{\bibinfo{volume}{66}},
  \bibinfo{pages}{041104} (\bibinfo{year}{2002}).

\bibitem[{\citenamefont{Arfken and Weber}(2001)}]{arfken}
\bibinfo{author}{\bibfnamefont{G.~B.} \bibnamefont{Arfken}} \bibnamefont{and}
  \bibinfo{author}{\bibfnamefont{H.~J.} \bibnamefont{Weber}},
  \emph{\bibinfo{title}{Mathematical Methods for Physicists}}
  (\bibinfo{publisher}{Harcourt Academic Press}, \bibinfo{address}{U.S.A.},
  \bibinfo{year}{2001}), \bibinfo{edition}{5th} ed.

\bibitem[{\citenamefont{Tatarskii}(1983)}]{tatarskii}
\bibinfo{author}{\bibfnamefont{V.~I.} \bibnamefont{Tatarskii}},
  \bibinfo{journal}{Sov. Phys. Usp.} \textbf{\bibinfo{volume}{26}},
  \bibinfo{pages}{311} (\bibinfo{year}{1983}).

\bibitem[{\citenamefont{Hillery et~al.}(1984)\citenamefont{Hillery, O'Connell,
  Scully, and Wigner}}]{wigner}
\bibinfo{author}{\bibfnamefont{M.}~\bibnamefont{Hillery}},
  \bibinfo{author}{\bibfnamefont{R.~F.} \bibnamefont{O'Connell}},
  \bibinfo{author}{\bibfnamefont{M.~O.} \bibnamefont{Scully}},
  \bibnamefont{and} \bibinfo{author}{\bibfnamefont{E.~P.}
  \bibnamefont{Wigner}}, \bibinfo{journal}{Phys. Rep.}
  \textbf{\bibinfo{volume}{106}}, \bibinfo{pages}{121} (\bibinfo{year}{1984}).

\bibitem[{\citenamefont{Serimaa et~al.}(1986)\citenamefont{Serimaa, Javanainen,
  and Varr\'o}}]{serimaa}
\bibinfo{author}{\bibfnamefont{O.~T.} \bibnamefont{Serimaa}},
  \bibinfo{author}{\bibfnamefont{J.}~\bibnamefont{Javanainen}},
  \bibnamefont{and} \bibinfo{author}{\bibfnamefont{S.}~\bibnamefont{Varr\'o}},
  \bibinfo{journal}{Phys. Rev. A} \textbf{\bibinfo{volume}{33}},
  \bibinfo{pages}{2913} (\bibinfo{year}{1986}).

\bibitem[{\citenamefont{Wilkie and Brumer}(1997{\natexlab{a}})}]{wilkie1}
\bibinfo{author}{\bibfnamefont{J.}~\bibnamefont{Wilkie}} \bibnamefont{and}
  \bibinfo{author}{\bibfnamefont{P.}~\bibnamefont{Brumer}},
  \bibinfo{journal}{Phys. Rev. A} \textbf{\bibinfo{volume}{55}},
  \bibinfo{pages}{27} (\bibinfo{year}{1997}{\natexlab{a}}).

\bibitem[{\citenamefont{Wilkie and Brumer}(1997{\natexlab{b}})}]{wilkie2}
\bibinfo{author}{\bibfnamefont{J.}~\bibnamefont{Wilkie}} \bibnamefont{and}
  \bibinfo{author}{\bibfnamefont{P.}~\bibnamefont{Brumer}},
  \bibinfo{journal}{Phys. Rev. A} \textbf{\bibinfo{volume}{55}},
  \bibinfo{pages}{43} (\bibinfo{year}{1997}{\natexlab{b}}).

\bibitem[{\citenamefont{Kling et~al.}(2006)\citenamefont{Kling, Siedschlag,
  Verhoef, Khan, Schultze, Uphues, Ni, Uiberacker, Drescher, Krausz
  et~al.}}]{kling}
\bibinfo{author}{\bibfnamefont{M.~F.} \bibnamefont{Kling}},
  \bibinfo{author}{\bibfnamefont{C.}~\bibnamefont{Siedschlag}},
  \bibinfo{author}{\bibfnamefont{A.~J.} \bibnamefont{Verhoef}},
  \bibinfo{author}{\bibfnamefont{J.~I.} \bibnamefont{Khan}},
  \bibinfo{author}{\bibfnamefont{M.}~\bibnamefont{Schultze}},
  \bibinfo{author}{\bibfnamefont{T.}~\bibnamefont{Uphues}},
  \bibinfo{author}{\bibfnamefont{Y.}~\bibnamefont{Ni}},
  \bibinfo{author}{\bibfnamefont{M.}~\bibnamefont{Uiberacker}},
  \bibinfo{author}{\bibfnamefont{M.}~\bibnamefont{Drescher}},
  \bibinfo{author}{\bibfnamefont{F.}~\bibnamefont{Krausz}},
  \bibnamefont{et~al.}, \bibinfo{journal}{Science}
  \textbf{\bibinfo{volume}{312}}, \bibinfo{pages}{246} (\bibinfo{year}{2006}).

\end{thebibliography}

\end{document}